# Broken Inversion Symmetry and Interface-induced Spin-polarization for metal-Weyl semimetal stacked interfaces

Tuhin Kumar Maji[1], Kumar Vaibhav[2], Samir Kumar Pal[1] and Debjani Karmakar[3,*]

[1]*Department of Chemical Biological and Macromolecular Sciences, S.N. Bose National Centre for Basics Sciences, Salt Lake, Sector 3, Kolkata 700106, India*

[2]*Computer Division, Bhabha Atomic Research Centre, Trombay, Mumbai 400085, India*

[3]*Technical Physics Division, Bhabha Atomic Research Centre, Trombay, Mumbai 400085, India*

*Corresponding Author: Debjani Karmakar: debjan@barc.gov.in

*Abstract*

Weyl semimetal TaAs, a congenial host to the massless Weyl fermions, spontaneously lacks the time-reversal and the inversion symmetry and thus effectuates topologically stable Weyl nodes, resembling magnetic monopoles in momentum space. Former experimental analysis had revealed that the near-zero spin-polarization of bulk TaAs experiences a boost in presence of point-contacts of non-magnetic metals along with the associated phenomena of tip-induced superconductivity, providing the impetus to study the large-area stacked interfaces of TaAs with Noble metals like Au and Ag. First-principles calculations on these interfacial systems have manifested an increment of the interface-induced spin-polarization and contact-induced transport spin-polarization. In contrast to the single interface, for stacked system, the broken inversion symmetry of the system introduces a z-directional band-dispersion resulting in an energetically separated series of Weyl cones. The Weyl cones for TaAs/Ag and TaAs/Au stacked interfaces are observed to be of type-I and type-II nature respectively. Thus, the current study demonstrates the designing of two different types of spin-polarized Weyl systems from non-magnetic metal and type I Weyl components.

INTRODUCTION:

Observation of massless Weyl fermion in topologically non-trivial Weyl semimetal systems like Tantalum and Niobium arsenides and phosphides[1,2] bridges the long-pending gap between its prediction in high-energy physics and the condensed matter realization as an eminent quasiparticle excitation[3]. Such observation complements the similar discovery of its Dirac[4,5] or Majorana[6,7] counterparts in periodic systems.

Crystalline Weyl semimetal (WSM) systems consist of bulk band crossings, named as Weyl nodes, consisting of non-degenerate three-dimensional bands near Fermi-level, in contrast to the two-dimensional band crossings at Dirac points of Graphene[8,9]. Near the nodal point, the low-energy physics is governed by Weyl equations[10], having solutions as massless Weyl spinors with distinct chirality, as obtained from relativistic field theory[3,11-14]. Realization of Weyl nodes is possible, when the time reversal symmetry (TRS) or inversion symmetry (IS) is broken for a Dirac semi-metal[4,5,15,16], where a Dirac node can be considered to be composed of two Weyl nodes of opposite chirality[17]. Thus, Weyl nodes in a crystalline system always appear in pairs of opposite chirality at two distinctly separated *k*-points, so that, their annihilation by simple translational symmetry preserving perturbations, are prevented, rendering the topological stability of the system[18]. The hallmark signature of WSM is the presence of Fermi-arcs on the surface, joining two Weyl nodes of opposite chirality [1,2,17,19,20]. The degeneracy associated with Weyl nodes at the 3D Weyl cones in the systems like $Cd_3As_2$ [5], $Na_3Bi$ [4,21], Ta and Nb phosphides and arsenides[1,2,17,19,20] or at the 2D surface states of topological insulators[6,8,22] depends merely on translational symmetry. Even before the discovery of Weyl systems, there was proposal of realization of 3D semimetal phase by alternate vertical stacks of topological insulators and band insulators[23].

A significant property of WSM is the singularity of Berry curvature (BC) at the Weyl nodes, representing a magnetic monopole in the momentum space with a distinct chirality [24-27]. These nodal points act as a source or a sink of the BC for the positive and negative chirality respectively and thus their pair-occurrence prevents the divergence of BC. Such topological systems are also defined by Chern numbers, representing the integral of BC over any closed 2D manifold at the Fermi-level[5,19]. The Chern number can have a nonzero (zero) value, depending on the inclusion (exclusion) of a Weyl node within the 2D manifold[20]. WSMs are well known to have anomalous nature of their DC transport related to the absence of backscattering and weak antilocalization[19,28,29], as a consequence of presence of topological surface states having well-defined spin. Another interesting chiral property of WSM is the chiral anomaly, where, in presence of electric and magnetic fields, the particle number corresponding to a particular valley is not conserved[29], culminating intervalley pumping of electrons between two nodes of opposite chirality[29-33].

In addition to the exciting fundamental physics, topological systems like $Cd_3As_2$[34] and TaAs[35-37] exhibited mesoscopic superconducting phase and high transport spin polarization in presence of metallic point contacts like Ag. However, regarding the occurrence of superconducting phase in TaAs, debates persist about the nature of the pairing symmetry. Whereas, Wang et al. had proposed unconventional *p*-wave superconductivity[37,38], Gayen et al. had demonstrated an *s*-wave conventional nature[36]. The unresolved controversies about the fundamental nature of superconductivity and the urge of understanding the interfacial effects has motivated the current investigation of large-area interfaces of Weyl semimetal system TaAs with two well-known noble metals Au and Ag.

In the next section, we brief the computational methodology used for the theoretical investigations. The successive section describes the structures and the underlying symmetry of the bare and stacked interfaces and their respective band structures. Phonon dispersion of the interfacial systems is described in the next section. The quantum transport properties of the devices with TaAs as channel material and Au/Ag as lateral/vertical contacts are described in the last section. The last section summarizes the obtained results with a conclusion.

**Computational Methodology**

The first principles electronic structure of the single and stacked TaAs/Au and TaAs/Ag interfaces are carried out by using the Vienna ab initio simulation package (VASP)[39,40] with norm-conserving projector augmented wave (PAW) pseudopotentials and generalized gradient approximated (GGA) Perdew-Burke-Ernzerhof (PBE) exchange-correlation functionals[41] with incorporation of spin-orbit (SO) coupling. For the pseudopotentials, the valence levels for Ta and As consist of $5p^66s^25d^3$ and $4s^24p^3$ configurations respectively.

Van der Waal corrections are incorporated by following the semi-empiricial Grimme DFT-D2 method[42]. The plane-wave cutoff and Monkhorst-Pack *k*-points grid[43] are set as 500 eV and 7×7×5 respectively. Ionic relaxations are performed by using conjugate gradient algorithm[44] with the cutoff for the Hellmann-Feynman force as 0.01 eV/Å.

For the density functional theory (DFT)-coupled quantum transport, we have used the Atomistic Toolkit 15.1 packages, with the GGA-PBE exchange correlation and double-zeta plus polarization (SZP) basis set. For each lateral/vertical interface, ionic optimizations are carried out to relax the interfaces with the real-space energy cutoff as 200 Hartree and the maximum force of 0.01eV/Å. For quantum transport calculations using the two-probe model, we have used the DFT coupled nonequilibrium Green's function (NEGF) method. Devices with appropriate channel length of TaAs, nullifying inter-electrode transmission and lateral/vertical contacts of Au/Ag, are constructed after fully relaxing the lateral and vertical interfaces with metals. The temperature of the calculation is set to 300K and 5K. At interfaces of the electrodes and the central region, Dirichlet boundary condition has been employed to ensure the charge neutrality in the source and the drain region. The channel length was optimized to ensure zero contributions from the inter-electrode transmissions. The Monkhorst-Pack *k*-point mesh is sampled with 5×5×50. The transport properties and the corresponding transmission coefficients are calculated by averaging over a $k_x \times k_y$ mesh of 10×10 in a direction perpendicular to the current transmission.

Γ-point centered transmission coefficients, perpendicular to the transport axis within the irreducible Brillouin zone (IBZ), are calculated for both the lateral and vertical device geometries by using $T^{\parallel}(E) = Tr[\Gamma_L^{\parallel}(E) \mathscr{G}_{\parallel}(E) \Gamma_R^{\parallel}(E) \mathscr{G}_{\parallel}^{\dagger}(E)]$, where, $\mathscr{G}_{\parallel}$ is the retarded Green's function, $\Gamma_{L/R}$ is the level broadening with respect to the corresponding self-energies of the electrodes. This function, after integrating over the *k*-point mesh in the IBZ, generates the transmission coefficient.

We have used supercell method to calculate the phonon dispersion and phonon DOS, where the dynamical matrices are calculated over a q-point grid 5×5×5, and the Hamiltonian derivatives are calculated over a k-mesh 10×10×10.

## Structural construction of Interfaces

Bulk TaAs belongs to a body centred (BC) tetragonal lattice structure with space group $I4_1md$ containing two formula units per unit cell. The system lacks an inversion symmetry and for its [001] surface, the $C_4$ rotational symmetry is also broken [1,2,17]. We have constructed the interfaces with the bilayer Au [111] and Ag[111] surfaces with the [3×3×2] supercell of TaAs. For the vertical interfaces TaAs/Au and TaAs/Ag, the mean interfacial strain is minimized to 1.4% after a mutual rotation between TaAs and metal [111] surface by ~ 29 degrees by following the Co-incident site lattice (CSL) method as implemented in ATK[45,46]. The underlying symmetry of the BC-tetragonal lattice transforms to a triclinic ($P_1$) one for the interface. Fig 1 depicts the structure of the supercell in $P_1$ symmetry, the corresponding interfaces and the Brillouin zone with the high-symmetry points for these systems. For TaAs/Au and TaAs/Ag single interface, we have added a vacuum of 15 Å in the z-direction to avoid the z-directional periodic replication. For stacked interfaces, the TaAs/metal stack repeats itself along z-direction without and vacuum-induced symmetry breaking.

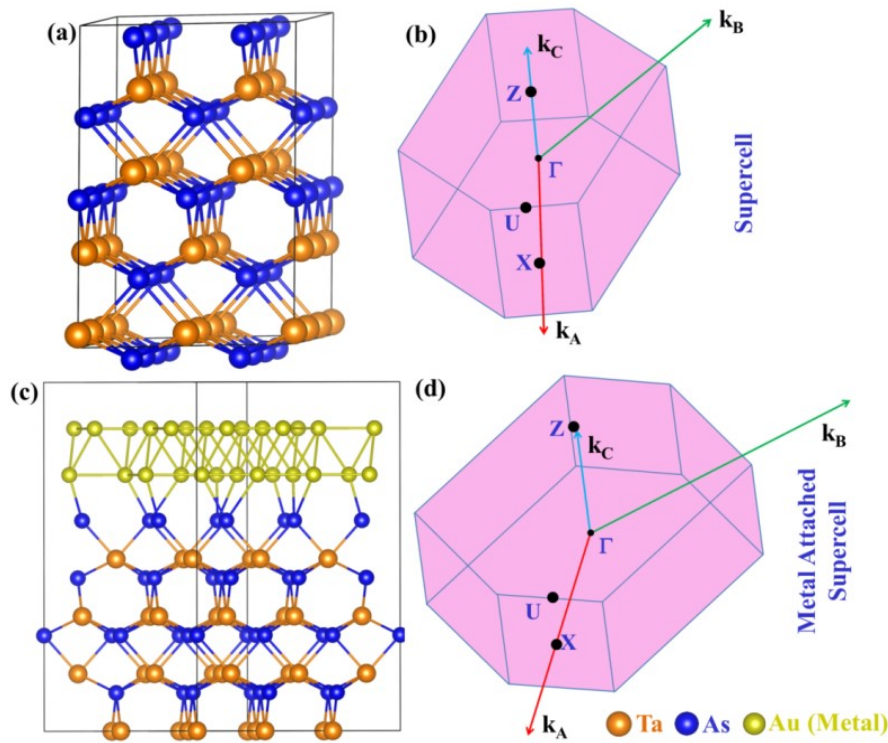

*Figure 1:* a) Structural image of TaAs supercell in $P^1$ symmetry, b)Brillouin Zone and high symmetry points of the supercell, c) structural image of TaAs/Au heterostructure, d)Brillouin Zone and high symmetry points of the heterostructure in $P^1$ symmetry.

## Electronic structure of TaAs for Triclinic symmetry

The orbital projected band structure of TaAs supercell (as depicted in Fig 1(a)) along high symmetry directions of the triclinic BZ (Fig 1(b)) is depicted in Fig 2(a). Along X-U and U-Z, there are non-degenerate 3D band-crossings, manifesting Weyl-cone like feature. Ta and As, being in the $3^+$ and $3^-$ valence states, have one-fifth filled Ta-$5d$ ($5d^2$) and one-third filled As-$4p$ valence levels contributing near $E_F$. Bands near $E_F$ have highly hybridized Ta-$5d$ and As-$4p$ character, as can be seen from the Fig 2(a). Fig 2(b) represents the orbital-projected density of states (DOS). Near $E_F$, the DOS shows the expected semi-metal like behavior[47]. Fig 2(c) shows the expected splitting of Weyl Cones after application of SO-coupling producing a fully gapped band structure[17,20].

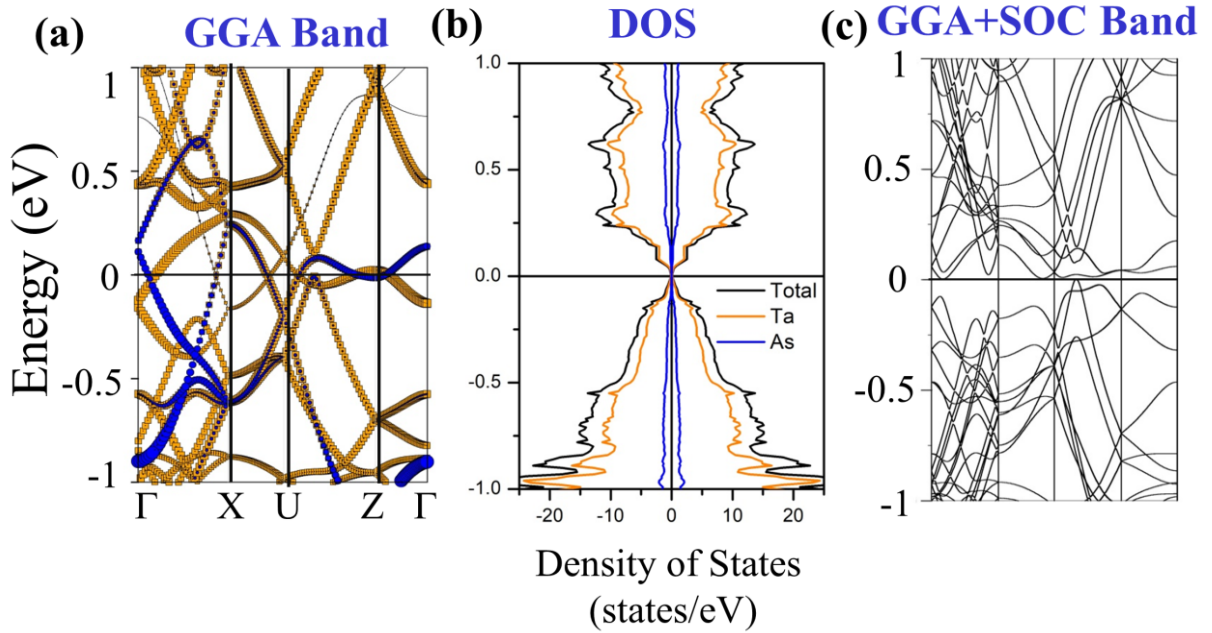

*Figure 2:* Orbital Projected a) band-structure and b) density of states of TaAs supercell using GGA, c) GGA+SO gapped band structure of the same supercell.

For TaAs/Au or TaAs/Ag single interfaces, presence of vacuum destroys the $z$-directional periodicity rendering 2D-like degenerate bands along X-U and Z-G, having almost zero z-dispersion as depicted in Fig 3(a). The DOS figures indicates three attributes as a result of the inter-layer charge transfer, *viz.* 1) destruction of the Weyl-cone feature, 2) doping of the TaAs layer underneath and 3) spin-polarization of the system as a whole. The mutual charge transfer with the metal layer introduces an *n*-type doping for both the TaAs/Au and TaAs/Ag systems. Pristine TaAs system has stabilized in AFM ground state, with zero resultant magnetic moment and near-zero magnetic moments for the constituent ions (Fig 2), implying its non-magnetic nature. Interface with metal introduces spin-polarization in the system due to the mutual transfer of carriers between the delocalized *s-d* hybridized metal layer and the TaAs layer. The ground state magnetic configuration for TaAs/Au and TaAs/Ag system is ferromagnetic (FM) and antiferromagnetic (AFM) respectively. Table I lists the magnetic moments, type of doping and ground state magnetic configuration of the interfacial systems.

Albeit introduction of spin-polarization, the broken inversion symmetry of the system is disturbed due to the presence of huge vacuum slab for the TaAs/Au and TaAs/Ag single interfaces, leading to the apparent destruction of Weyl feature. The Weyl cone and the corresponding formation of Weyl node like band crossing can however be restored upon formation of z-stacked interfaces like TaAs/Au/TaAs/Au… and TaAs/Ag/TaAs/Ag…. Fig 4 represents the atom-projected band-structures and DOS for the stacked interfaces. The most fascinating effect of stacked interfaces is that the band crossings along X-U have retrieved back the Weyl node feature for both the systems, as can be seen from Fig 4. For TaAs/Ag, the Weyl cone is having type I nature[48,49], whereas, for TaAs/Au, it is tilted with respect to $E_F$, manifesting type II [48,49] behaviour. In addition, the Weyl crossings along X-U are distributed for the overall energy range, such that, by appropriate doping induced shift of the Fermi-level with the help of applied bias, the system is capable of retaining its Weyl nodal features. The doping nature and the magnetic ground states have undergone a change with respect to the single interfaces, with the TaAs/Ag stacked interface having a *p*-type doping and AFM configuration of the spin-magnetic moments, as can be seen from Table I. The hole-doped antiferromagnetism in TaAs/Ag system renders it to be a potential strongly correlated system like the cuprates and pnictides [50,51]. The band structure and DOS, as presented in Fig 4, displays a strong mixing of metal *s-d* hybridized bands and Ta-5*d* and As-4*p* levels. The charge-density and spin-density plots, as presented in Fig 5, also support the fact of increase of spin-polarization for stacked interfaces with a significantly visible charge and spin-density for otherwise invisible densities for pristine system.

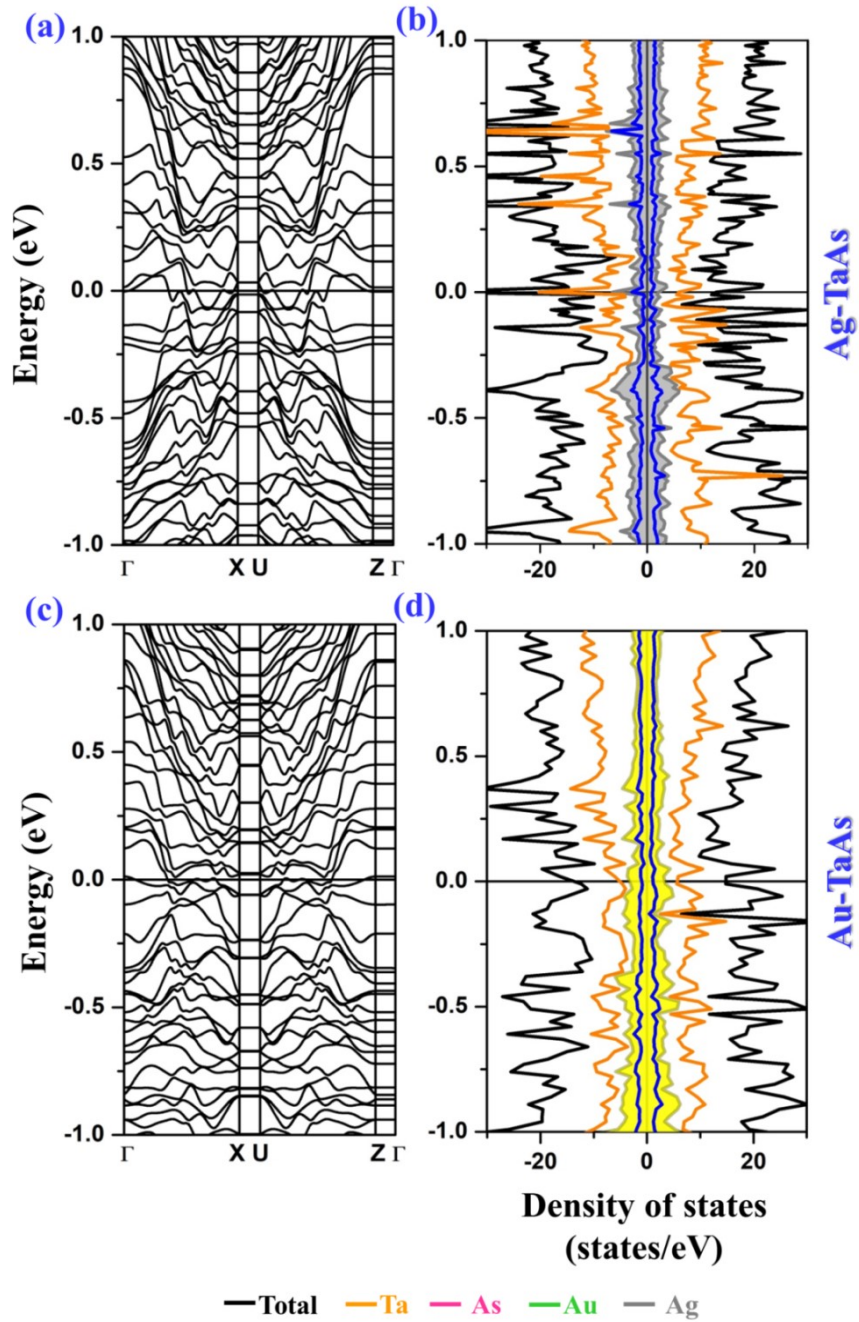

*Figure 3:* a) Band structure and the corresponding b) orbital Projected DOS of Ag-TaAs single interface with vacuum and c) band structure and corresponding d) orbital Projected DOS of Au-TaAs single interface with vacuum using GGA.

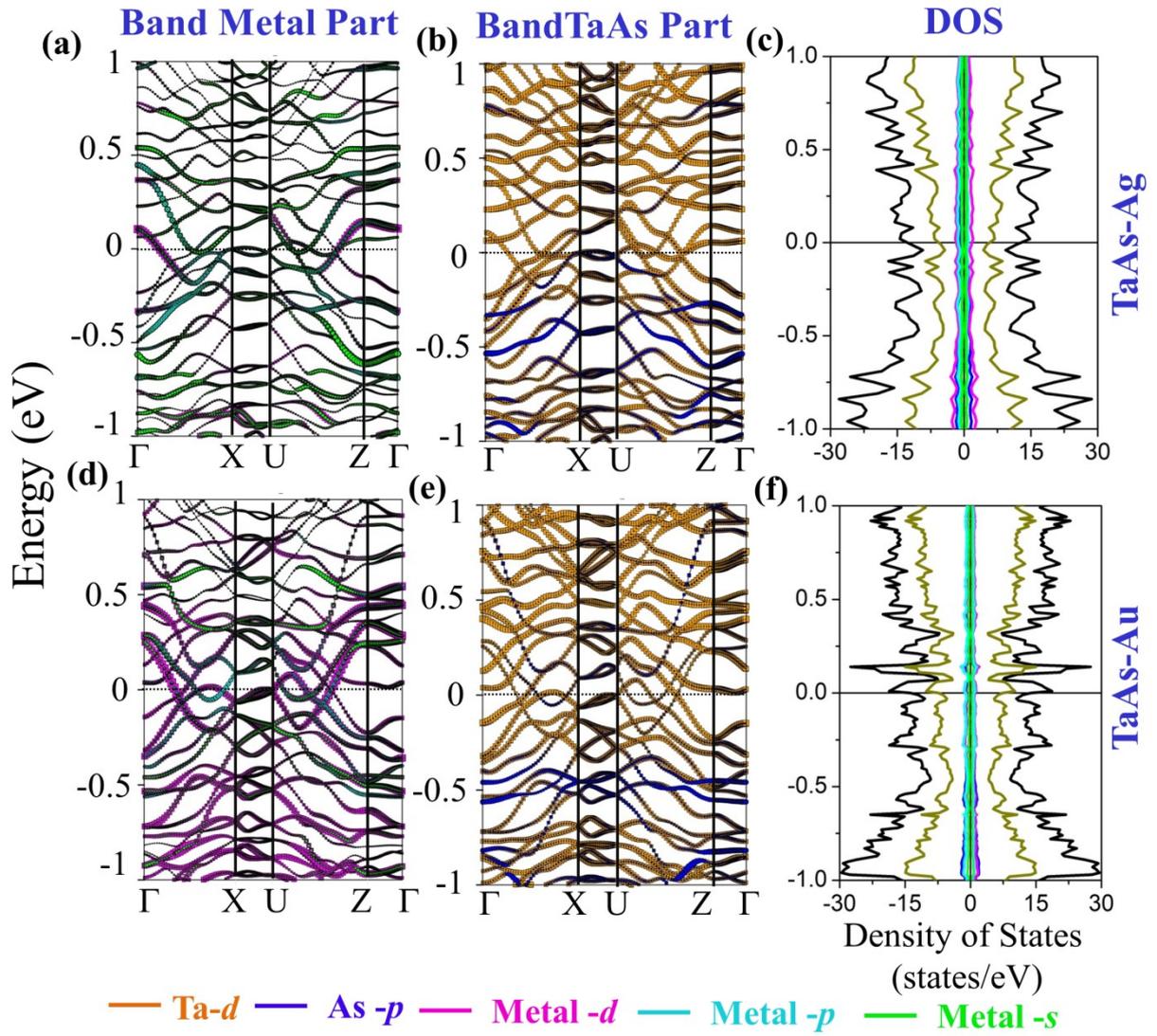

*Figure 4:* Orbital Projected 'FAT-Band' of different systems a) TaAs part of Ag-TaAs, c) Ag part of Ag-TaAs, and c) corresponding DOS. Orbital Projected 'FAT-Band' of d) TaAs part of Au-TaAs, c) Au part of Au-TaAs, and c) corresponding DOS.

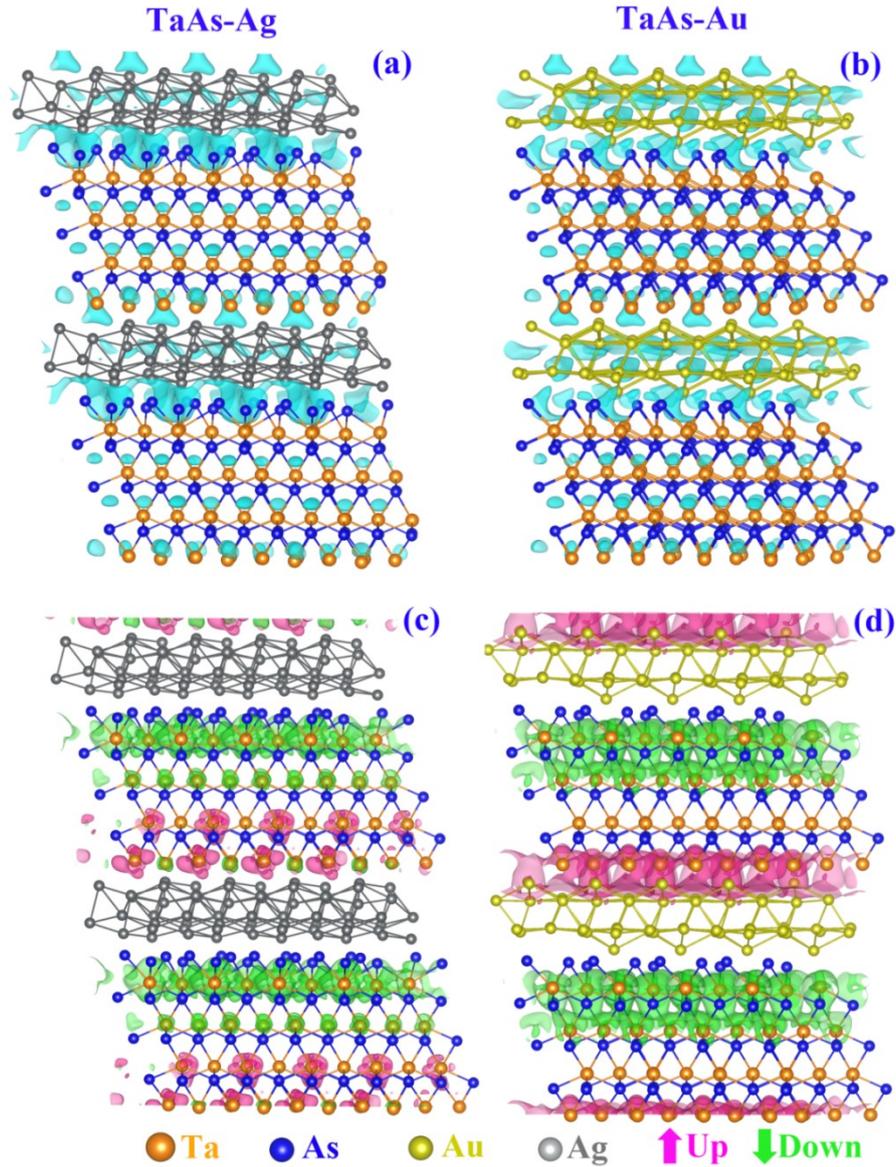

*Figure 5:* *Charge density plot of stacked interfaces a) Ag-TaAs and b) Au-TaAs and spin density plot of c) Ag-TaAs and d) Au-TaAs.*

**Phonon dispersion for stacked interfaces**

In order to obtain the nature of phonon dispersion of TaAs in presence of Au or Ag metal layer, we have created a smaller stacked interface as shown in first column of Fig 6. Corresponding to each structure, the phonon band dispersion and the DOS are plotted in the same row. The band dispersion for bulk TaAs resembles with the literature[52,53], where the well-dispersed acoustic modes, having the most contribution from the vibration of the heavier Ta-ions are separated from the more localized optical counterpart due to vibrations of the lighter As-ions by a band gap ~ 2.4 eV. Occurrence of phonon band-gaps due to the mass-discrepancy of the contributing ions is well-known[54]. In presence of stacked interfaces with metals, the band-gap disappears with a significant increase of density of bands and levels mostly in the acoustic frequency range, as can be seen from the second and third column of Fig 6. The increase in density of acoustic phonon bands indicates an increase of electron-phonon interactions for the stacked interfaces TaAs/Au and TaAs/Ag.

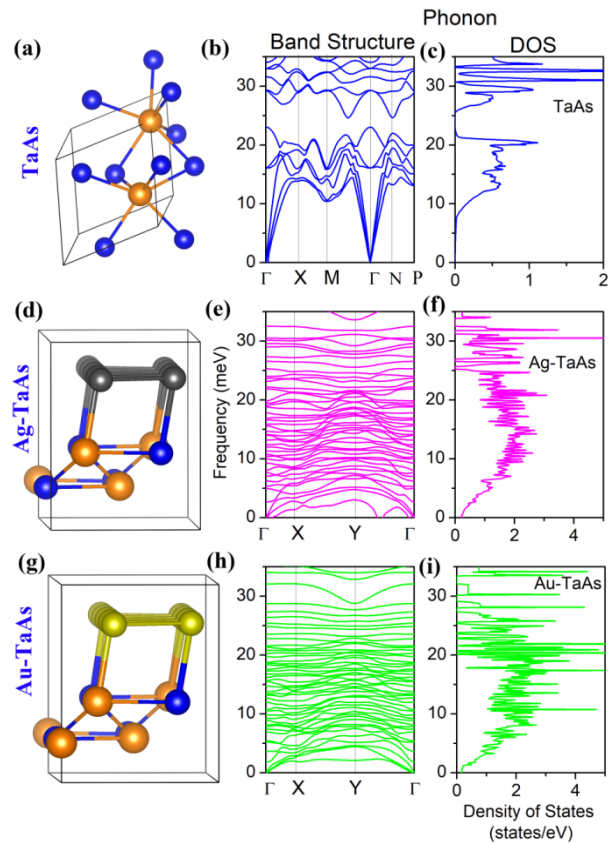

*Figure 6:* a) Structural representation of TaAs b) corresponding phonon band structure and c) DOS of TaAs. D) Structural representation of Ag-TaAs e) corresponding phonon band structure and f) DOS of Ag-TaAs. G) Structural representation of Au-TaAs b) corresponding phonon band structure and c) DOS of Au-TaAs.

Thus, in presence of metals, there is an eloquent impact on the phonon dispersion and densities of TaAs, suggesting emergence of phonon-induced correlated behaviour within the system.

**Quantum transport for TaAs devices with Au/Ag contacts**

As seen by the point-contact spectroscopic measurements, the transport spin-polarization of TaAs undergoes an increase in presence of Au or Ag tip contacts[34]. In this section, we have investigated the transport properties of two probe devices made out of TaAs as channel and Au/Ag as lateral or vertical contacts. The schematic constructions of lateral and vertical contact devices are shown in Fig 7(a) and (b).

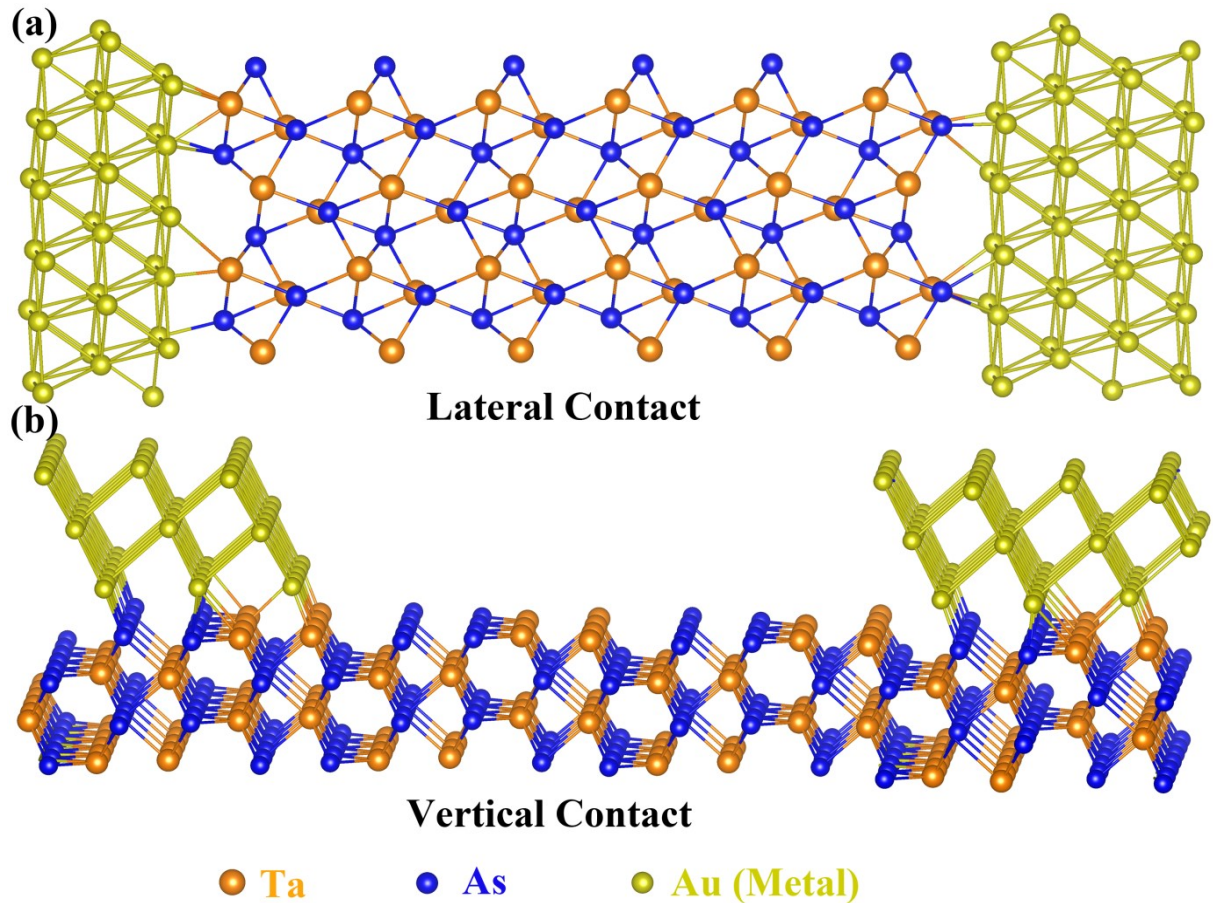

*Figure 7:* a) Structural representation of a) TaAs with lateral metal contact, and b) TaAs with vertical metal contact.

Fig. 8 represents the current ($I$) versus voltage ($V$) characteristics for the lateral and vertical contact systems with Au and Ag contacts at room temperature. In room temperature transport characteristics, transport spin-polarization is more prominent for lateral contacts. In Fig 9, we have plotted a comparison of the percentage of transport spin polarization for different bias voltages, calculated from its absolute value $P_t = (I_\uparrow - I_\downarrow)/(I_\uparrow + I_\downarrow)$ for 5K and 300K for both lateral and vertical contact geometry. On an average, for both systems, the value of $P_t$ is an order of magnitude more for lateral contacts with a trend of increase in polarization after decreasing temperature. The increase of transport spin polarization in presence of metal contacts was also observed in prior experimental studies [34]. For vertical contacts, on the other hand, $P_t$ at room temperature transport is more. Lateral contacts, as an outcome of perfect covalent bonding and thereby obtained smother transport, are more effective in retaining the spin of the injected electrons at the interface.

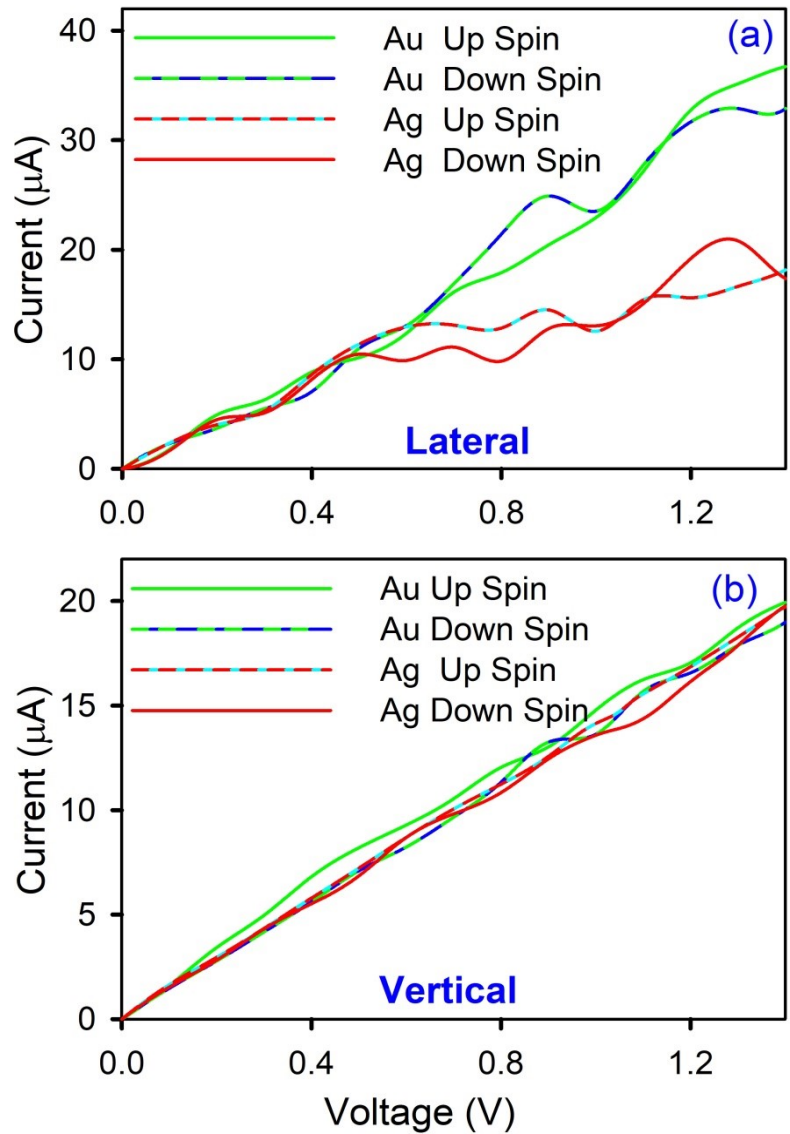

**Figure 8:** *Transport characteristics for a) lateral contacts b) vertical contact geometries for both TaAs-Au and TaAs-Ag.*

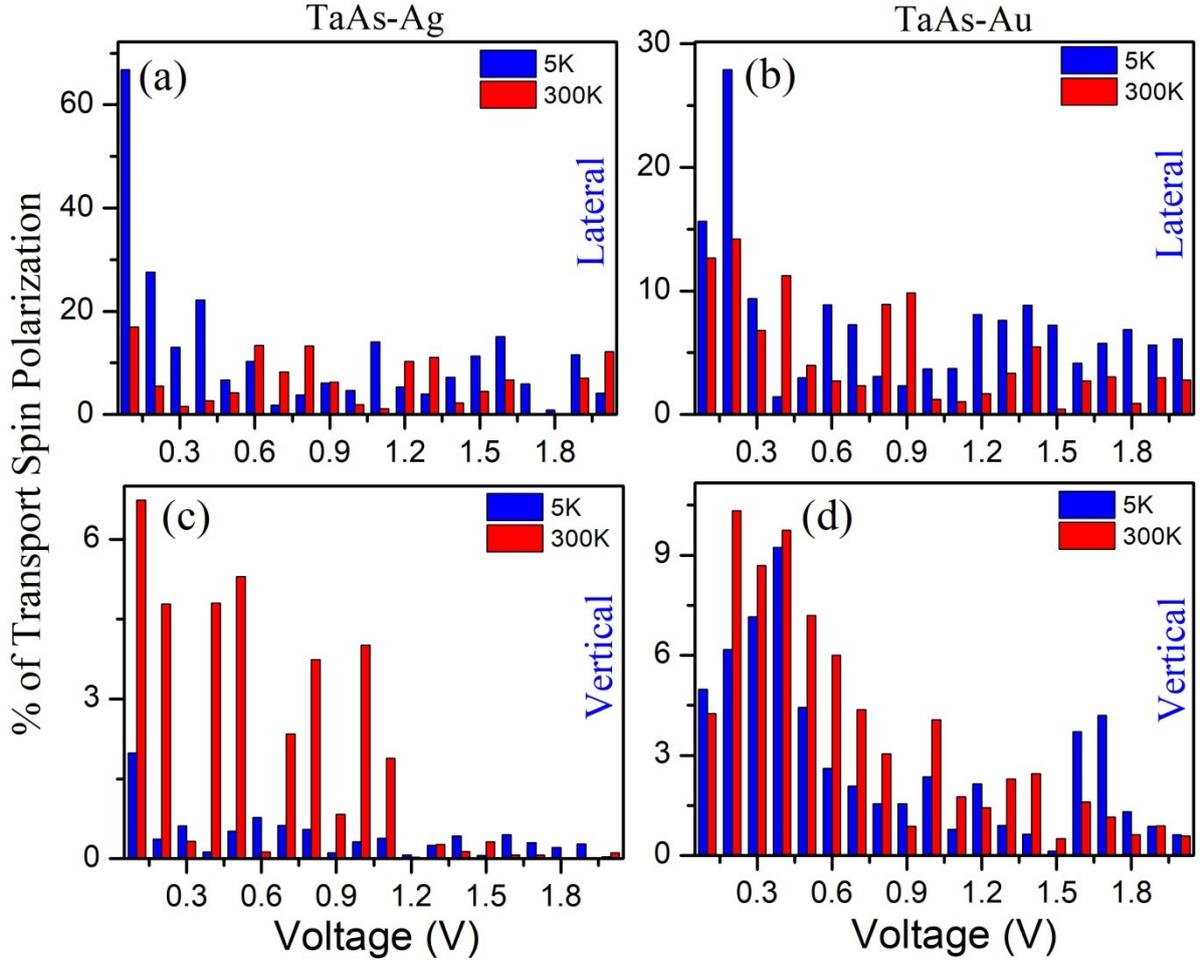

*Figure 9:* *Percentage of transport spin polarization of a) TaAs-Ag lateral, b) TaAs-Au lateral, c)TaAs-Ag vertical and d) TaAs-Au vertical contact geometries.*

Fig 10 presents the total (up-spin + down-spin) transmission spectra and the corresponding interpolated color-maps at 1V on the Γ- centred $K_A$- $K_B$ plane *perpendicular to the transport axis*. As anticipated for the lateral contacts, they have shown a uniform transmission across the transport axis, whereas the vertical ones are having various transmission zones, varying from higher to lower transmission. The average transmission is higher in Ag-contacts for lateral devices, whereas it is almost similar for the vertical contacts for both Ag and Au.

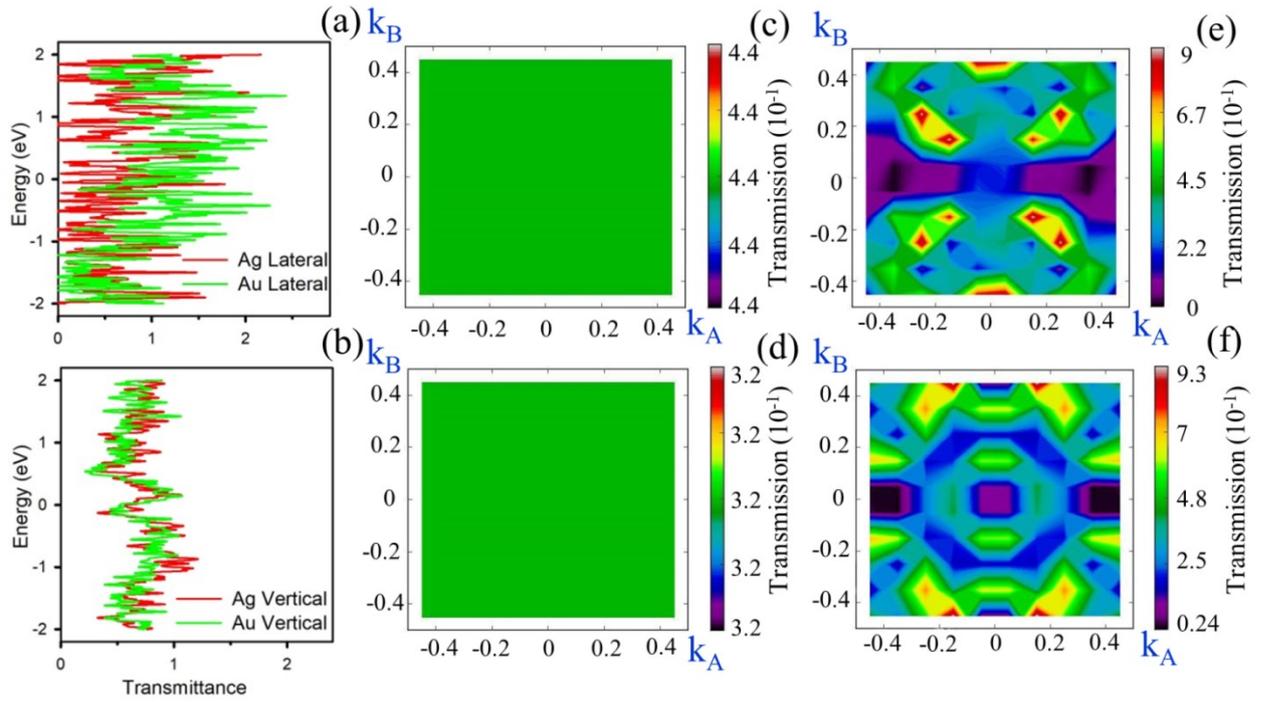

*Figure 10:* *Transmittance of a) Lateral Ag-TaAs and Au-TaAs, b) vertical Ag-TaAs and Au-TaAs, Transmission colour map of c) Ag-Lateral, d) Au- Lateral, e) Ag- vertical and f) Au-vertical.*

The local density of states (LDOS), presented in Fig 11, depict the difference of transport between the two types of devices *along the device transport axis*. Whereas, the lateral contacts show less interfacial scattering at contacts, implying smoother transmission, the vertical contacts have more scattering near contact boundary. LDOS features complement with the obtained larger transport spin polarizations for lateral contacts, where lesser interfacial scattering promotes an effective spin transport.

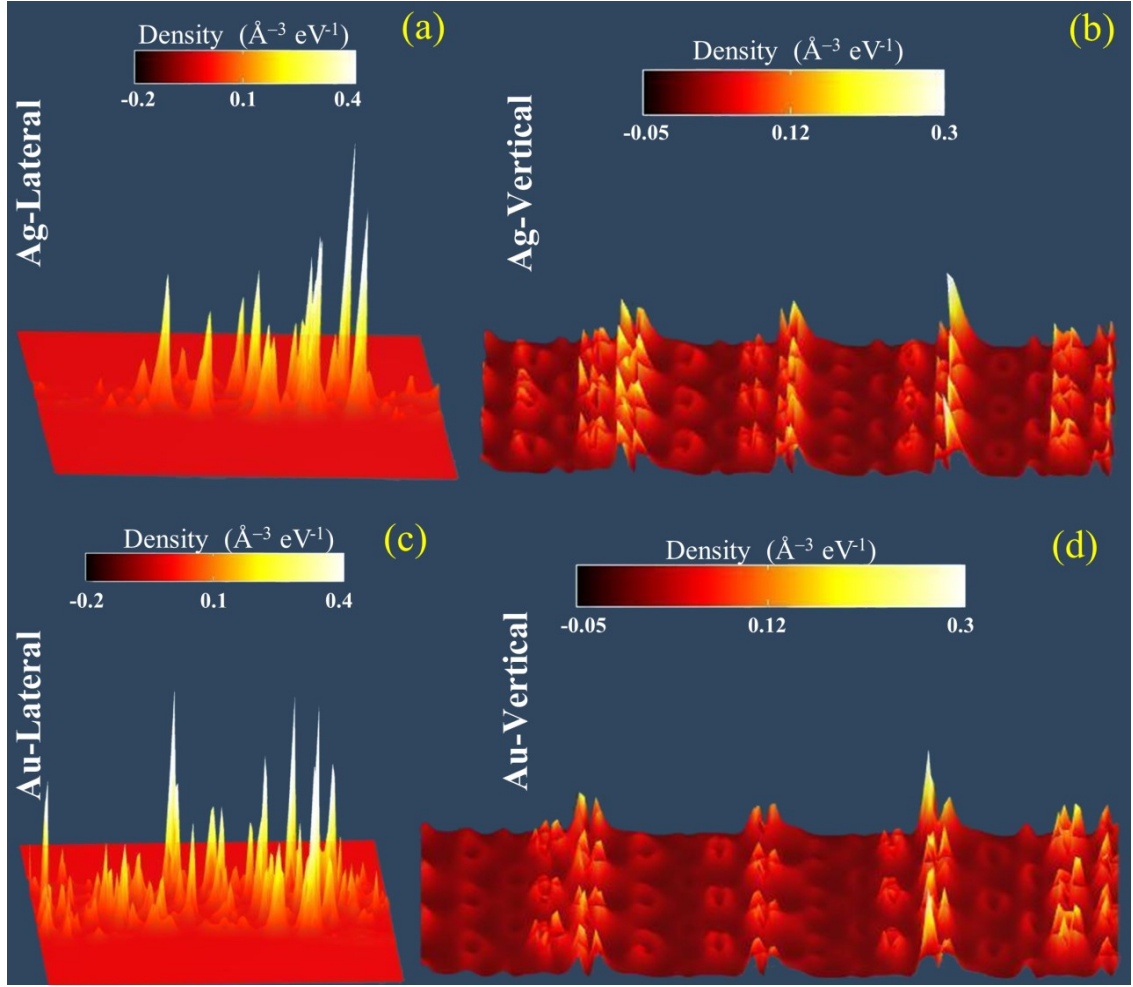

*Figure 11:* *Local density of states (LDOS) for a) Ag-Lateral, b) Ag- vertical, c) Au-Lateral and d) Au- vertical contact devices.*

Thus, the types of contact metal and the geometry of contacts have tremendous impact on spin-polarized device transport of devices with large-area interface.

## Conclusion

In summary, we have investigated the macroscopic single and stacked interfaces of the Weyl semimetal system TaAs with two noble metals Au and Ag. The band-structures for stacked interfaces have manifested the impact of inversion symmetry breaking and Type I and II Weyl cones for Ag and Au systems respectively. Whereas, pristine TaAs is non-magnetic, TaAs/Au and TaAs/Ag show electron-doped FM and hole doped AFM ground-states. The phonon-dispersion for the stacked systems shows a disappearing phonon band gap and an increase of acoustic phonon density of states. Transport properties of lateral and vertical metal contact devices have significant impact of transport spin-polarization at room temperature and at low temperature. Lateral contacts, on an average, have more uniform transport and more transport spin-polarization with lowering of temperature.

**Table 1:** Magnetic moments, nature of doping and ground state magnetic configurations of single interface and stacked interface systems.

|  | System | Magnetic Moment ($\mu_B$) | Nature of doping | Ground State Configuration |
|---|---|---|---|---|
| Single Interface | TaAs | 0.7520 | - | FM |
|  | Ag-TaAs | 2.6760 | *n*-type | AFM |
|  | Au-TaAs | 3.9020 | *n*-type | FM |
| Stacked Interface | TaAs | 0 | - | Non-Magnetic |
|  | Ag-TaAs | -0.0013 | *p*-type | AFM |
|  | Au-TaAs | -0.0081 | *n*-type | FM |

# Acknowledgement

TKM wishes to acknowledge the support of DST India for INSPIRE Research Fellowship and SNBNCBS for funding. We also thank DAE (India) for financial grant 2013/37P/73/BRNS. DK would like to acknowledge BARC ANUPAM supercomputing facility for computational resources.